\documentclass[aps,prl,showpcs,amsmath,amssymb,longbibliography,twocolumn,showpacs,notitlepage]{revtex4-1}
\usepackage{multirow}
\usepackage{epsfig}
\usepackage{amsmath}
\usepackage{bm}
\usepackage{times}
\usepackage{graphicx}
\usepackage{color}
\usepackage{slashed}
\usepackage{graphicx}
\usepackage{amsmath} 
\usepackage{tikz}
\usetikzlibrary{positioning,shapes}
\usepackage{relsize} 
\usepackage[latin1]{inputenc}
\usepackage{hyperref}  

\def\bea{\begin{eqnarray}}
\def\eea{\end{eqnarray}}
\def\bean{\begin{equation*}}
\def\eean{\end{equation*}}

\begin{document} 
 
\title{Dark Matter Interpretation of the Neutron Decay Anomaly}

\author{Bartosz~Fornal}
\affiliation{Department of Physics, University of California, San Diego, 9500 Gilman Drive, La Jolla, CA 92093, USA}
\author{Benjam\'{i}n~Grinstein}
\affiliation{Department of Physics, University of California, San Diego, 9500 Gilman Drive, La Jolla, CA 92093, USA} 
\date{\today}

\begin{abstract}
There is a long-standing discrepancy between the neutron lifetime measured in beam and bottle experiments. 
We propose to explain this anomaly by a dark decay channel for the neutron, involving one or more dark sector particles in the final state. If any of these particles are stable, they can be the dark matter. We construct representative particle physics models  consistent with all experimental constraints. \vspace{8mm}
\end{abstract}

\maketitle

\section{Introduction}

The neutron is one of the fundamental building blocks of matter. Along with the proton and electron it makes up most of the visible universe. Without it, complex atomic nuclei simply would not have formed.  Although the neutron was discovered over eighty years ago \cite{1932RSPSA.136..692C} and has been studied intensively thereafter, its precise lifetime is still an open question \cite{2011RvMP...83.1173W,Greene}.
The dominant neutron decay mode is $\beta$ decay

\vspace{-6mm}
\bea
n \rightarrow p + e^- \!+ \bar\nu_e \ ,\nonumber
\eea
\vspace{-6mm}

\noindent
described by the matrix element
\bea
\mathcal{M} = \tfrac{1}{\sqrt2}\,{G_F V_{ud}\, g_V}\left[ \,\bar{p} \,\gamma_\mu n - \lambda \,\bar{p}\, \gamma_5\gamma_\mu n\, \right] \left[ \, \bar{e} \,\gamma^\mu (1-\gamma_5) \nu \,\right]. \nonumber
\eea
The theoretical estimate for the neutron lifetime is \cite{Wilkinson:1982hu,Czarnecki:2004cw,Marciano:2005eca,Hardy:2014qxa}

\vspace{-5mm}
\bea
\tau_n = \frac{4908.7(1.9)\,{\rm s}}{|V_{ud}|^2(1+3\,\lambda^2)} \ .\nonumber
\eea
The Particle Data Group (PDG) world average for the axial-vector  to vector coupling ratio is $\lambda = -1.2723\pm0.0023$ \cite{Patrignani:2016xqp}.
Adopting the PDG average $|V_{ud}| = 0.97417 \pm 0.00021$  
gives $\tau_n$  between $875.3\, {\rm s}$ and $891.2 \, {\rm s}$ within $3 \, \sigma$.

There are two qualitatively different types of direct neutron lifetime measurements: bottle and beam experiments. 

In the first method, ultracold neutrons are stored in a container for a time comparable to the neutron lifetime. The remaining neutrons that did not decay are counted and fit to a decaying exponential, $\exp(-t/\tau_n)$. The average from the five bottle experiments included in the PDG \cite{Patrignani:2016xqp} world average  is \cite{Mampe,Serebrov:2004zf,Pichlmaier:2010zz,Steyerl:2012zz,Arzumanov:2015tea}

\vspace{-9mm}
\bea
\tau_n^{\rm bottle}  = 879.6 \pm 0.6  \ {\rm s} \ .\nonumber
\eea
\vspace{-5mm}

\noindent
Recent measurements using trapping techniques \cite{Serebrov:2017bzo,Pattie:2017vsj} yield a neutron lifetime within $2.0\,\sigma$ of this average.

In the beam method, both the number of neutrons $N$ in a beam and the protons resulting from $\beta$ decays are counted, and the lifetime is obtained from the  decay rate, $dN/dt = -N/\tau_n$. This yields a considerably longer neutron lifetime; the average from the two beam experiments included in the PDG average \cite{Byrne:1996zz,Yue:2013qrc} is
\bea
\tau_n^{\rm beam}  = 888.0 \pm 2.0  \ {\rm s} \ .  \nonumber
\eea

The discrepancy between the two results is  $4.0 \, \sigma$. This suggests that either one of the measurement methods suffers from an uncontrolled systematic error, or there is a theoretical reason why the two methods give different results.

In this paper we focus on the latter possibility. We assume that the discrepancy between the neutron lifetime measurements arises from an incomplete theoretical description of neutron decay and we investigate how the Standard Model (SM)  can be extended to account for  the anomaly.

\section{Neutron dark decay}

Since in beam experiments neutron decay is observed by detecting
decay protons, the lifetime they measure  is related to the actual neutron lifetime by
\bea
\tau^{\rm beam}_n = \frac{\tau_n}{{\rm Br}(n\rightarrow p + {\rm anything})} \ .
\eea
In the SM the branching fraction (Br), dominated by $\beta$ decay,  is $100\%$ and the two lifetimes are the same.
The neutron decay rate obtained from bottle experiments is
\bea
\Gamma_n  \simeq 7.5 \times 10^{-28} \ {\rm GeV}. \nonumber
\eea
The discrepancy $\Delta \tau_n \simeq 8.4 \ {\rm s}$ between the values measured in bottle and beam experiments corresponds to \footnote{It was speculated in \cite{Greene} that the neutron lifetime discrepancy might be caused by hypothetical oscillations of neutrons into mirror neutrons \cite{Berezhiani:2005hv}.}
\bea
\Delta \Gamma_n^{\rm exp} = \Gamma_n^{\rm bottle} - \Gamma_n^{\rm beam} \simeq 7.1 \times 10^{-30} \ {\rm GeV} \ .\nonumber
\eea 

We propose that this difference be explained by the existence of a {\emph{dark decay}} channel for the neutron, which makes 
\bea
{\rm Br}(n\rightarrow p + {\rm anything}) \approx 99\% \ .\nonumber
\eea
There are two qualitatively different scenarios for the new dark decay channel, depending on whether the final state consists entirely of dark particles or contains visible ones:
\bea
&&{\rm (a)} \ \  n \rightarrow \, {\rm invisible} \,+\, {\rm visible} \ ,\nonumber\\ [3pt]
&& {\rm (b)} \ \  n \rightarrow \, {\rm invisible} \ .\nonumber
\eea
Here the label ``invisible'' includes dark sector particles, as well as neutrinos.
Such decays are described by an effective operator
$\mathcal{O} =X n$,
where $n$ is the neutron and $X$ is a spin $1/2$ operator, possibly composite, e.g.~$X=\chi_1 \chi_2 ... \chi_{k}$, with the $\chi$'s being fermions and bosons combining into spin $1/2$. 
From an experimental point of view, channel ${\rm (a)}$ offers a detection possibility, whereas channel ${\rm (b)}$ relies on higher order radiative processes. We provide examples of both below.

\noindent
{\emph{Proton decay constraints}}\\
The operator $\mathcal{O}$ generally gives rise to proton decay via 
\bea
p \rightarrow n^* + e^+ + \nu_e \ ,\nonumber
\eea
followed by the decay of $n^*$ through the channel ${\rm (a)}$ or ${\rm (b)}$  and has to be suppressed \cite{Miura:2016krn}. 
Proton decay can be eliminated from the theory if the sum of masses of particles in the minimal final state $f$ of neutron decay, say $M_{f}$, is larger than $m_p-m_e$. On the other hand, for the neutron to decay, $M_f$ must be smaller than the neutron mass, therefore it is required that

\vspace{-8mm}
\bea
m_p-m_e < M_f < m_n \ .\nonumber
\eea

\vspace{2mm}
 
 \noindent
{\emph{Nuclear physics bounds}}\\
In general, the decay channels ${\rm (a)}$ and ${\rm (b)}$ could trigger nuclear transitions from $(Z,A)$ to $(Z,A-1)$. If such a transition is accompanied by a prompt emission of a state $f'$ with the sum of masses of particles making up $f'$ equal to $M_{f'}$, it can  be eliminated from the theory by imposing
$
M_{f'} >  \Delta M = M(Z,A) - M(Z,A-1) \ .\nonumber
$
Of course $M_{f'}$ need not be the same as $M_f$, since the final state $f'$ in nuclear decay may not be available in neutron decay. For example, $M_{f'} < M_f$ when the state $f'$ consists of a single particle, which is not an allowed final state of the neutron decay. If $f'\!=\!f$ then $f'$ must contain at least two particles. 
The requirement becomes, therefore,

\vspace{-7mm}
\bea
\Delta M <  {\rm min} \big\{M_{f'}\!\big\} \leq  M_f  < m_n\ . \nonumber
 \eea
The most stringent of such nuclear decay constraints comes from the requirement of $^9{\rm Be}$ stability, for which  $\Delta M = 937.900 \ {\rm MeV}$, thus
\bea\label{constr}
937.900 \ {\rm MeV} <  {\rm min} \big\{M_{f'}\!\big\} \leq M_f <  939.565 \ {\rm MeV} \ . \ \ \ \ \ \
\eea
The condition in Eq.\,(\ref{constr}) circumvents  all nuclear decay limits listed in PDG \cite{Patrignani:2016xqp}, including the most severe ones  \cite{Ahmed:2003sy,Araki:2005jt,Takhistov:2015fao}. 

\vspace{3mm}
\noindent
{\emph{Dark matter}}\\
Consider $f$ to be a two-particle final state containing a dark sector spin $1/2$ particle $\chi$. Assuming the presence of the interaction ${\chi}\,n$, the condition in Eq.\,(\ref{constr}) implies that the other particle in $f$ has to be a photon or a dark sector particle $\phi$ with mass $m_\phi < 1.665 \ {\rm MeV}$ (we take it to be spinless). 
The decay $\chi \rightarrow p + e^- \!+ \bar{\nu}_e$ is forbidden if
\bea\label{DMpd}
m_\chi < m_p + m_e = 938.783 \ {\rm MeV}\ .
\eea
Provided there are no other decay channels for $\chi$, Eq.\,(\ref{DMpd}) ensures that $\chi$ is stable, thus making it a DM candidate. 
On the other hand, if $\chi \rightarrow p + e^- \!+ \bar{\nu}_e$ is allowed, although this prevents $\chi$ from being the DM, its lifetime is still long enough to explain the neutron decay anomaly. In both scenarios $\phi$ can be a DM particle as well.

Without the interaction $\chi  \,n$, only the sum of  final state masses is constrained by Eq.\,(\ref{constr}). 
Both $\chi$ and $\phi$ can be DM candidates, provided 
\bea
|m_\chi - m_\phi| < m_p + m_e \ . \nonumber
\eea
One can also have a scalar DM particle $\phi$ with mass $ m_\phi < 938.783 \ {\rm MeV}$ and $\chi$ being a Dirac right-handed neutrino. Trivial model-building variations are implicit. 
The scenarios with a Majorana fermion $\chi$ or a real scalar $\phi$ are additionally constrained by neutron-antineutron oscillation and dinucleon decay searches \cite{Abe:2011ky,Gustafson:2015qyo}. 

\section{Model-independent analysis}

Based on the discussed experimental constraints, the available channels for the neutron dark decay  are:  
\bea
 n\rightarrow \chi\,\gamma\ , \  \ \ \ n\rightarrow \chi\,\phi \ , \ \ \ \ n\rightarrow \chi\,e^+e^- \ ,   \nonumber
 \eea
 as well as those involving additional dark particle(s) and/or photon(s). 
 \vspace{-2mm}

\subsection{Neutron \ $\rightarrow$ \ dark matter  \ + \   photon}

\vspace{-1mm}
This decay is realized in the case of a two-particle interaction involving the fermion DM $\chi$ and a three-particle interaction including $\chi$ and a photon, i.e.,
$
 \chi \,n \,,  \, \chi\, n\, \gamma
$.
Equations (\ref{constr}) and (\ref{DMpd}) imply that the DM mass is
\bea
937.900 \ {\rm MeV} <  m_\chi <  938.783 \ {\rm MeV} \nonumber
\eea
 and the final state photon energy
 \bea\label{phE}
 0.782 \ {\rm MeV} <E_\gamma < 1.664 \ {\rm MeV} \ .
 \eea
We are not aware of any experimental constraints on such monochromatic photons.
The search described in \cite{Nico:2006pe,Cooper:2010zza,Bales:2016iyh} measured photons from radiative $\beta$ decays in a neutron beam, however,  photons were recorded only if they appeared in coincidence with a proton and an electron, which is not the case in our proposal.
 
To describe the decay $n\rightarrow \chi\,\gamma$ in a quantitative way, we consider theories with an interaction $\chi\,n$, and an interaction $ \chi\, n\, \gamma$  mediated by a mixing between the neutron and $\chi$. An example of such a theory is given by the effective Lagrangian
\bea\label{F0}
\mathcal{L}^{\rm eff}_1 &=& \bar{n}\,\big(i\slashed\partial-m_n +\tfrac{g_ne}{2 m_n}\sigma^{\mu\nu}F_{\mu\nu}\big) \,n \nonumber\\
&&+ \   \bar{\chi}\left(i\slashed\partial-m_\chi\right) \chi + \varepsilon \left(\bar{n}\chi + \bar{\chi}n\right) \ ,
\eea
where $g_n \!\simeq\! -3.826$ is the neutron $g$-factor and $\varepsilon$ is the mixing parameter with dimension of mass. The  term corresponding to $n\rightarrow \chi\,\gamma$ is obtained by transforming Eq.\,(\ref{F0}) to the mass eigenstate basis and, for $\varepsilon \ll m_n-m_\chi$,  yields
\bea\label{eff1c0}
\mathcal{L}_{n \rightarrow \chi \gamma}^{\rm eff} =\frac{g_n e}{2m_n}\frac{\varepsilon}{(m_n-m_\chi)}\,\bar\chi \,\sigma^{\mu\nu} F_{\mu\nu} \,n \ .
\eea
Therefore, the neutron dark decay rate is
\bea\label{rate-phase000}
\Delta\Gamma_{n\rightarrow \chi\gamma} &=& \frac{g_n^2e^2}{8\pi}\bigg(1-\frac{m_\chi^2}{m_n^2}\bigg)^3  \frac{m_n\,\varepsilon^2}{(m_n-m_\chi)^2} \nonumber\\
&\approx&  \Delta \Gamma_{n}^{\rm exp} \,\big(\tfrac{1+x}{2}\big)^3 \Big(\tfrac{1-x}{1.8\times 10^{-3}}\Big)\Big(\tfrac{\varepsilon \ [{\rm GeV}]}{9.3\times 10^{-14}}\Big)^2\!, \ \ \ \ \ \ \
\eea
where $x=m_\chi/m_n$. The rate is maximized when $m_\chi$ saturates the lower bound $m_\chi = 937.9 \ {\rm MeV}$. A particle physics realization of this case is provided by model 1 below.

The testable prediction of this class of models is a  monochromatic photon with an energy in the range specified by Eq.\,(\ref{phE}) and a branching fraction
\bea
\frac{\Delta\Gamma_{n\rightarrow \chi\gamma}}{\Gamma_n} \approx 1\% \ .\nonumber
\eea
A signature involving an  $e^+e^-$ pair with total energy $E_{e^+e^-} < 1.665 \ {\rm MeV}$ is also expected, but with a  suppressed branching fraction of at most $1.1\times  10^{-6}$.

If $\chi$ is not a DM particle,  the bound in Eq.\,(\ref{DMpd}) no longer applies and the final state monochromatic photon can have an energy in a wider range:
\bea
0 <E_\gamma < 1.664 \ {\rm MeV} \ ,
\eea
entirely escaping detection as $E_\gamma \rightarrow 0$.

\subsection{Neutron \  $\rightarrow$ \  two dark particles}

\vspace{-1mm}
Denoting  the final state dark fermion and scalar by $\chi$ and $\phi$, respectively, and an intermediate dark fermion by $\tilde\chi$, consider a scenario with both a two- and three-particle interaction,
$
{{\tilde\chi}} \,n \,, \,  {{\chi}}\, n \,\phi  .
$
The requirement in Eq.\,(\ref{constr}) takes the form
\bea\label{DMmasss}
937.900 \ {\rm MeV} < m_\chi + m_\phi <  939.565 \ {\rm MeV} \ ,
\eea
and both $\chi$, $\phi$ are stable if 
\bea
|m_\chi - m_\phi| < 938.783 \ {\rm MeV} \ . \nonumber
\eea
Also,
 $m_{\tilde\chi} > 937.900 \ {\rm MeV}$.
 
If $m_{\tilde\chi}>m_n$, the only  neutron dark decay channels are $n\rightarrow \chi\, \phi$ and $n\rightarrow \tilde\chi^* \rightarrow p + e^- \!+\bar{\nu}_e$, with branching fractions governed by the strength of the ${{\chi}}\, n \,\phi$ interaction.   Even if this coupling is zero, the lifetime of $\tilde\chi$  is long enough for  the anomaly to be explained.
In the case $937.9 \ {\rm MeV} < m_{\tilde\chi}< m_n$, the particle  $\tilde\chi$  can be produced on-shell and there are three neutron dark decay channels: $n\rightarrow \tilde\chi\, \gamma$,  $n\rightarrow \chi\, \phi$ and $n\rightarrow \tilde\chi^* \rightarrow p + e^- \!+ \bar{\nu}_e$ (when $m_{\tilde\chi}>938.783 \ {\rm MeV}$), with branching fractions depending on the strength of the ${{\chi}}\, n \,\phi $ coupling. The rate for the decay $n\rightarrow \tilde\chi^* \rightarrow p + e^- \!+ \bar{\nu}_e$ is negligible compared to that for $n\rightarrow \tilde\chi\, \gamma$. In the limit of a vanishing ${{\chi}}\, n \,\phi $ coupling  this case reduces to $n\rightarrow \tilde\chi \,\gamma$.

An example of such a theory is
\bea\label{F3}
\mathcal{L}^{\rm eff}_{2} &=& \mathcal{L}^{\rm eff}_{1}(\chi \rightarrow \tilde\chi) +  \left(\lambda_\phi \,\bar{\tilde{\chi}}\, \chi\, \phi + {\rm h.c.}\right)\nonumber\\
&&+ \  \bar{\chi}\left(i\slashed\partial-m_{\chi}\right) \chi + \partial_\mu \phi^* \partial^\mu \phi - m_\phi^2 |\phi|^2 \ .
\eea
The term corresponding to $n\rightarrow \chi \,\phi$ is
\bea\label{eff1c2}
\mathcal{L}_{n \rightarrow \chi \phi}^{\rm eff} =\frac{\lambda_\phi\,\varepsilon}{m_n-m_{\tilde\chi}} \,\bar{\chi}  \,n \,\phi^* \ .
\eea
This yields the neutron dark decay rate
\bea\label{rate-phase}
\Delta\Gamma_{n\rightarrow \chi\phi} = \frac{|\lambda_\phi|^2}{16\pi}\sqrt{f(x, y)}\, \frac{m_n\,\varepsilon^2}{(m_n-m_{{\tilde\chi}})^2} \ ,
\eea
where

\vspace{-7mm}
\bea
f(x, y) =[(1-x)^2-y^2] \, [(1+x)^2-y^2]^3 \nonumber
\eea
\vspace{-6mm}

\noindent
with
$x=m_\chi/m_n$ and $y=m_\phi/m_n$.
A particle physics realization of this scenario is provided by model 2 below.

For $m_{\tilde\chi} > m_n$ the missing energy signature has a branching fraction $\approx 1 \%$. There  will also be a very suppressed radiative process involving a photon in the final state with a branching fraction $3.5 \times 10^{-10}$ or smaller.

As discussed earlier, in the case $937.9 \ {\rm MeV} < m_{\tilde\chi} < m_n$ both the visible and invisible neutron dark decay channels are present. The ratio of their branching fractions is
\bea\label{competition}
\frac{\Delta\Gamma_{n\rightarrow \tilde\chi\gamma}}{\Delta\Gamma_{n\rightarrow \chi\phi}} = \frac{2g_n^2e^2}{|\lambda_\phi|^2} \frac{(1-\tilde{x}^2)^3 }{\sqrt{f(x, y)}}\ ,
\eea
where $\tilde{x}=m_{\tilde\chi}/m_n$, while their sum accounts for the neutron decay anomaly, i.e.
\bea
\frac{\Delta\Gamma_{n\rightarrow \tilde\chi\gamma} + \Delta\Gamma_{n\rightarrow \chi\phi}}{\Gamma_n} \approx 1\% \ .\nonumber
\eea
The branching fraction for the process involving a photon in the final state ranges from $\sim 0$  to $ 1\%$, depending on the masses and couplings.  A suppressed decay channel involving $e^+e^-$ is also present.

\subsection{Neutron \ $\rightarrow$ \  dark matter  \ + \   $e^+e^-$}

\vspace{-1mm}
This case is realized when the four-particle effective interaction involving the neutron, DM and an $e^+e^-$ pair is present and 
${\rm Br}(n\rightarrow \chi\,e^+e^-) \approx 1\%$. 
The requirement on the DM mass from Eq.\,(\ref{constr}) is
\bea
937.900 \ {\rm MeV} <  m_\chi <  938.543 \ {\rm MeV}  \nonumber
\eea
 and the allowed energy range of the $e^+e^-$ pair is
 \bea
2\,m_e \leq E_{e^+e^-}< 1.665 \ {\rm MeV} \ .\nonumber
 \eea
Assuming the effective  term for $n \rightarrow \chi \,e^+e^-$ of the form
\bea\label{eff1c0}
\mathcal{L}_{n \rightarrow \chi e^+e^-}^{\rm eff} =\kappa\,\bar\chi \,n \, \bar e\, e  \nonumber
\eea
 and a suppressed 
two-particle interaction $\chi \,n$, the neutron dark decay rate is
\bea\label{rate-phase00}
\Delta\Gamma_n &=& \frac{\kappa^2m_n^5}{128\, \pi^3} \int_{4z^2}^{(1-x)^2} \hspace{-1mm }\frac{d\xi}{\sqrt\xi} \left(\xi - 4z^2\right)^{\frac32}\left[(1+x)^2-\xi\right]\nonumber\\
&&\times \ \sqrt{(1-x^2-\xi)^2-4\,\xi \,x^2}\ ,\nonumber
\eea
where $x=m_\chi/m_n$ and $z=m_e/m_n$. It is maximized for $m_\chi = 937.9 \ {\rm MeV}$, in which case it requires $1/\sqrt\kappa \approx 670\  {\rm GeV}$ to explain the anomaly. We will not analyze further this possibility, but we note that a theory described by the Lagrangian (\ref{F3}) with $\phi$ coupled to an $e^+e^-$ pair could be an example.

\section{Particle physics models}

We now present two microscopic renormalizable models that are representative of the cases  $n\rightarrow \chi\,\gamma$ and $n\rightarrow \chi\,\phi$.

\subsection{Model 1}

\begin{figure}[t!]
\includegraphics[width=0.6\linewidth]{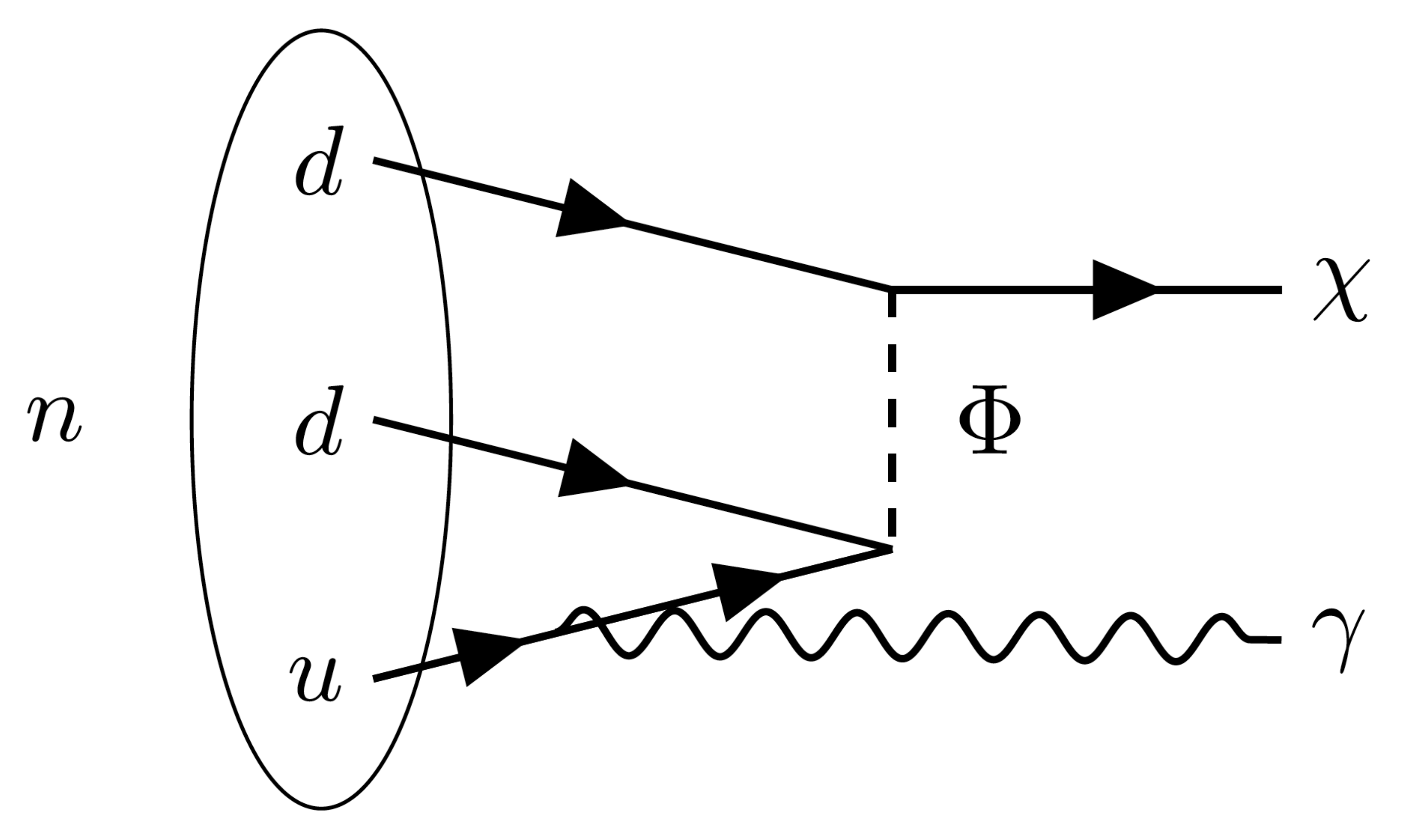}  
\caption{\small{Dark decay of the neutron in model 1.}}\vspace{0mm} 
\label{fig1}
\end{figure}

\vspace{-3mm}
The minimal model for the neutron dark decay requires only two particles beyond the SM: a scalar $\Phi=(3,1)_{-1/3}$ (color triplet, weak singlet, hypercharge $-1/3$), and a Dirac fermion $\chi$ (SM singlet, which can be the DM). 
This model is a realization of the case $n\rightarrow \chi\,\gamma$. The neutron dark decay proceeds through the process shown in Fig.\,\ref{fig1}.    
The   Lagrangian of the model is
\bea
\mathcal{L}_{1} &=&   \big(\lambda_q \,\epsilon^{ijk}\, \overline{u^c_L}_{i}\, d_{Rj} \Phi_k + \lambda_\chi\Phi^{*i}\bar\chi \,d_{Ri} +  \lambda_l \,\overline{Q^c_R}_{i}\, l_{L} \Phi^{*i}\nonumber\\
 & +& \! \lambda_Q \,\epsilon^{ijk}\, \overline{Q^c_R}_{i}Q_{Lj} \Phi_k + {\rm h.c.}\big) - M_\Phi^2 |\Phi|^2 - m_\chi \,\bar\chi\,\chi  \ , \ \ \ \ \ \ \ \ 
\eea
where $u^c_L$ is the complex conjugate of $u_R$. We assign baryon numbers $B_\chi\!=\!1$, $B_\Phi=-2/3$ and, to forbid proton decay \cite{Arnold:2012sd,Dorsner:2016wpm,Assad:2017iib}, assume baryon number conservation, i.e. set $\lambda_l = 0$ \footnote{The assumption of a small $\lambda_l$ is not necessary in the framework of the recently constructed grand unified theory with no proton decay \cite{Fornal:2017xcjd}}. For simplicity, we choose $\lambda_Q=0$.
The rate for $n \!\rightarrow \!\chi \gamma$ is given by Eq.\,(\ref{rate-phase000}) with

\vspace{-6mm}
\bea
\varepsilon = \frac{\beta\,\lambda_q\lambda_\chi }{M_{\Phi}^2} \nonumber
\eea

\vspace{-3mm}
\noindent
and $\beta$ defined by
\bea\langle 0| \epsilon^{ijk} \!\left(\overline{u^c_L}_{i} d_{Rj}\right) \!d_{Rk}^\rho |n\rangle = \beta \, \left(\tfrac{1+\gamma_5}{2}\right)^{\rho}_{\ \sigma} \, u^\sigma \ .\nonumber
\eea
Here $u$ is the neutron spinor, $\sigma$ is the spinor index and the parenthesis denote spinor contraction.
Lattice QCD calculations give $\beta = 0.0144(3)(21) \
  {\rm GeV}^3$ \cite{Aoki:2017puj},  where the errors are
    statistical and systematic, respectively.
Assuming $m_\chi= 937.9 \ {\rm MeV}$ to maximize the rate, the parameter choice explaining the anomaly is
\bea\label{rrate}
\frac{|\lambda_q\lambda_\chi|}{M_\Phi^2}  \approx 6.7 \times 10^{-6}  \  {\rm TeV}^{-2} \  .
\eea

In addition to the monochromatic photon with energy $E_\gamma < 1.664 \ {\rm MeV}$ and the $e^+e^-$ signal, one may search 
directly also for  $\Phi$.  It can be singly produced through $p \,p \rightarrow \Phi$ or pair produced via gluon fusion $g\, g \rightarrow \Phi\, \Phi$. This results in a dijet or four-jet signal  from $\Phi \rightarrow d^c u^c$, as well as a monojet plus missing energy signal from $\Phi \rightarrow d \,\chi$. Given Eq.\,(\ref{rrate}), $\Phi$ is not excluded by recent LHC analyses \cite{Aaboud:2017yvp,Sirunyan:2016iap,Aaboud:2017nmi,Khachatryan:2014lpa,Aaboud:2017phn,Sirunyan:2017hci} provided $M_\Phi \gtrsim 1 \ {\rm TeV}$ \footnote{A similar model with a scalar $\Phi = (3,1)_{2/3}$ and a Dirac fermion $\chi$ would also work. Since the scalar $(3,1)_{2/3}$ cannot couple to two first generation quarks, the rate in Eq.\,(\ref{rrate}) would be suppressed by the strange quark content of the neutron and would require a larger value of $|\lambda_q\lambda_\chi|/M_\Phi^2$.  Another viable option for $\Phi$ is the vector $(3,2)_{1/6}$.}.

If $\chi$ is a DM particle, without an efficient annihilation channel one has to invoke non-thermal DM production to explain its current abundance. This can be realized via a late decay of a new heavy scalar, as shown in \cite{Allahverdi:2017edd} for a similar model. Current DM direct detection experiments provide no constraints \cite{Angloher:2015ewa}.

The parameter choice in Eq.\,(\ref{rrate}) is excluded if $\chi$ is a Majorana particle, as in the model proposed in \cite{McKeen:2015cuz}, by the neutron-antineutron oscillation and dinucleon decay constraints \cite{Abe:2011ky,Gustafson:2015qyo}.  Neutron decays considered in \cite{Davoudiasl:2014gfa} are too suppressed to account for the neutron decay anomaly. 

\subsection{Model 2}

\begin{figure}[t!]
\includegraphics[width=0.77\linewidth]{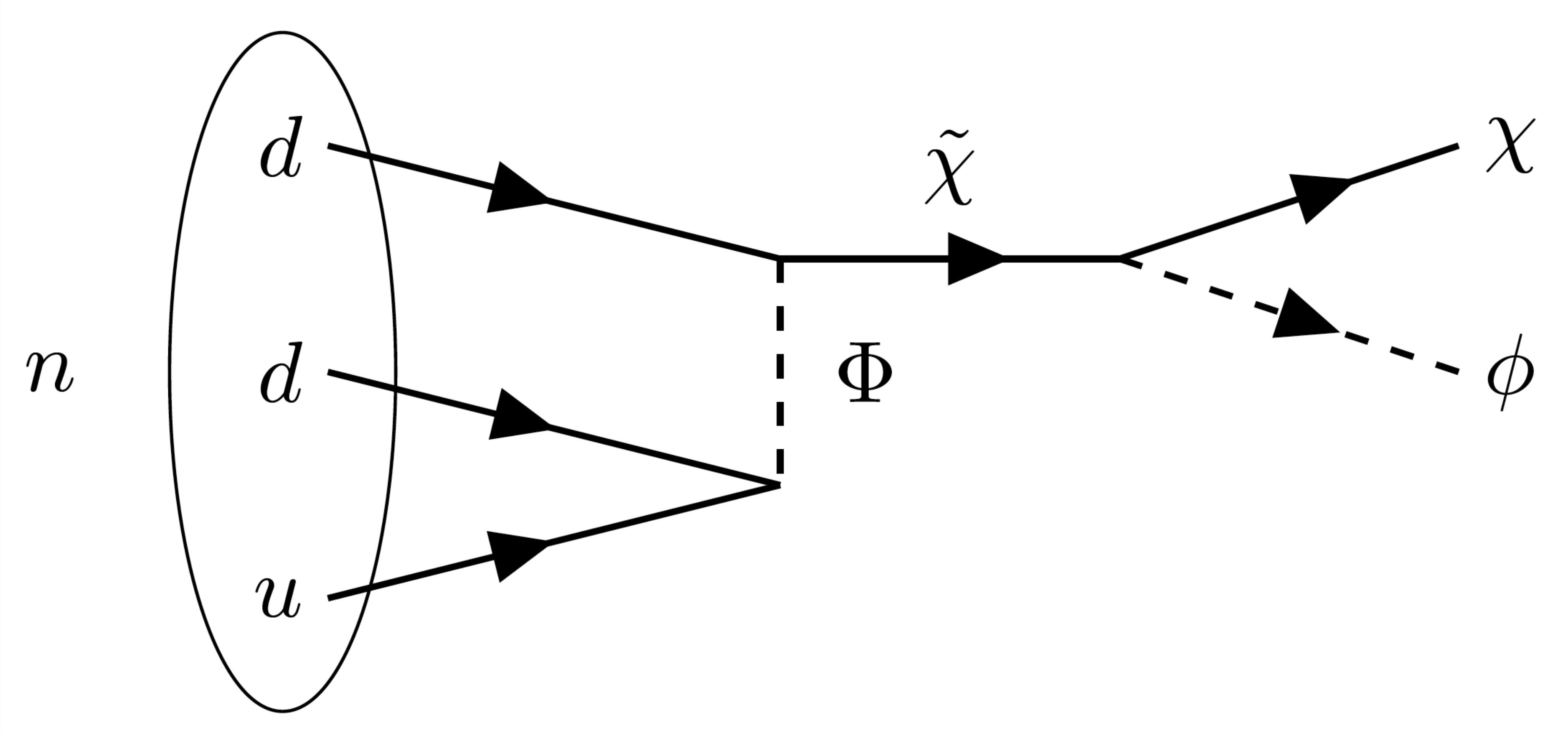}\vspace{-0.3mm}
\caption{\small{Dark decay of the neutron in model 2.}}\vspace{0mm}
\label{fig2}
\end{figure}

\vspace{-3mm}
A representative model for the case $n\rightarrow \chi\,\phi$ involves four new particles: the scalar $\Phi=(3,1)_{-1/3}$, two Dirac fermions $\tilde\chi$, $\chi$, and a complex scalar $\phi$, the last three being SM singlets. 
The neutron dark decay in this model is shown in Fig.\,\ref{fig2}.    
The Lagrangian is
\bea
\mathcal{L}_{2}  &=& \mathcal{L}_{1}(\chi \rightarrow \tilde\chi)   +  ( \lambda_\phi  \,\bar{\tilde\chi}\, \chi \,\phi  + {\rm h.c.})  -  m_\phi^2 |\phi|^2  -  m_\chi \,\bar\chi\,\chi   \ . \nonumber\\
\eea

\vspace{-2mm}
\noindent
Assigning $B_{\tilde\chi} = B_\phi = 1$ and $B_{\chi}=0$, baryon number is conserved. We have also imposed an additional $U(1)$ symmetry under which $\chi$ and $\phi$ have opposite charges.
For $m_\chi > m_\phi$ the annihilation channel  $\chi \,\bar\chi \rightarrow \phi\,\bar\phi$ via a $t$-channel $\tilde\chi$ exchange is open. The observed  DM relic density, 
assuming $m_{\chi} =937.9 \ {\rm MeV}$ and $m_\phi \approx 0$,
is obtained  for  $\lambda_\phi \simeq 0.037$. Alternatively, the DM can be non-thermally produced.

The rate for $n \rightarrow \chi \, \phi$  is described by Eq.\,(\ref{rate-phase}) with $\varepsilon = {\beta\,\lambda_q\lambda_\chi }/{M_{\Phi}^2}$. 
For $m_{\tilde\chi} \!=\! m_\chi$, the anomaly is explained with
\bea
\frac{|\lambda_q\lambda_\chi|}{M_\Phi^2}\frac{|\lambda_\phi|}{0.04} \approx  4.9 \times 10^{-7} \ {\rm TeV^{-2}}. \nonumber
\eea
For $\lambda_\phi \approx 0.04$ this is consistent with  LHC searches, provided again that $M_\Phi \gtrsim 1 \ {\rm TeV}$. Direct DM detection searches present no constraints.
For similar reasons as before, $\chi$ and $\tilde\chi$ cannot be Majorana particles.   

As discussed above, in this model the branching fractions for the visible (including a photon) and invisible final states can be comparable, and their relative size is described by Eq.\,(\ref{competition}). A final state containing an $e^+e^-$ pair is also possible. The same LHC signatures are expected as in model 1.

\section{Conclusions}

The puzzling discrepancy between the neutron lifetime measurements has persisted for over twenty years.~We could not find any theoretical model for this anomaly in the literature. In this paper we bring the neutron enigma into attention by showing that it  can be explained by a neutron dark decay channel with an unobservable particle in the final state.   Our proposal is phenomenological in its nature and the simple particle physics models provided serve only as an illustration of selected scenarios.

Despite most of the energy from the neutron dark decay escaping into the dark sector, our proposal is experimentally  verifiable. The most striking signature is monochromatic photons with energies less than 1.664 MeV. Furthermore, if the dark particle is the dark matter, the energy of the photon is bounded  by 0.782 MeV from below. The simplest model predicts the neutron decay into dark matter and a photon with a branching fraction of approximately $1 \%$. 
Another signature consists of  electron-positron pairs with total energy less than 1.665 MeV.  
It would be interesting to perform a detailed analysis of  the experimental reach for such signals.

Evidence for neutron dark decay can also be searched for
  in nuclear processes. There are several unstable isotopes with a
  neutron binding energy $S(n) <1.665 \ {\rm MeV}$ and a sufficiently
  long lifetime to probe the dark decay channel when the dark particle
  mass $m_\chi < m_n - S(n)$ \cite{Wang}. Consider, for example, $^{11}{\rm Li}$, for which $S(n) =
    0.396 \ {\rm MeV}$. $^{11}{\rm Li}$ $\beta$ decays with a lifetime 8.75 ms.
However, in the presence of a dark particle $\chi$ the decay
chain\,
$
^{11}{\rm Li} \rightarrow \, ^{10}{\rm Li} +
\chi \rightarrow \, ^9{\rm Li} + n + \chi 
$\,
becomes
available. $^9{\rm Li}$'s long lifetime, 178.3~ms, can be used to
  discriminate against background from $^{11}{\rm Li}$ $\beta$
  decay. A possible background comes from  $^9{\rm Li}$ production in
  $\beta$-delayed  deuteron emission from $^{11}{\rm Li}$ \cite{KELLEY201288,Raabe:2008rj}. 

From a theoretical particle physics perspective, our analysis  opens
the door to rich model building opportunities well beyond the two
simple examples we provided, including multi-particle dark
  sectors. Perhaps the dark matter mass being close to the nucleon
mass can explain the matter-antimatter asymmetry of the universe via a
similar mechanism as in asymmetric dark matter models. One may
  also include dark matter self-interactions without spoiling the general features of our proposal, used to address for example the core-cusp problem \cite{Spergel:1999mh}.

Finally,  the neutron lifetime has profound consequences for nuclear physics and astrophysics, e.g., it affects the primordial helium production during nucleosynthesis \cite{Mathews:2004kc} and impacts the neutrino effective number determined from the cosmic microwave background \cite{Capparelli:2017tyx}. 
If the neutron dark decay channel we propose is the true explanation for the difference in the results of bottle and beam experiments, then the correct value for the neutron lifetime is $\tau_n \simeq 880 \ {\rm s}$.  \vspace{7mm}

This research was supported in part by the DOE Grant No.~${\rm DE}$-${\rm SC0009919}$.

\noindent
{\bf\textit{Note added:}} \\
Inspired by the proposal in this paper, several experimental and theoretical efforts have already been undertaken.

In \cite{Tang:2018eln} the scenario $n \rightarrow \chi \gamma$ has been challenged experimentally for  $0.782 \ {\rm MeV} <E_\gamma < 1.664 \ {\rm MeV}$. The case $E_\gamma < 0.782 \ {\rm MeV} $ remains unexplored.

In \cite{Sun:2018yaw} the case ${\rm Br}(n \rightarrow \chi\,e^+e^-) = 1\%$ has been excluded for $E_{e^+e^-} \gtrsim 2 \,m_e + 100 \ {\rm keV}$.

In \cite{McKeen:2018xwc,Baym:2018ljz,Motta:2018rxp} the implications for neutron stars have been considered. 
For dark matter with self-interactions, the neutron dark decays in a neutron star can be sufficiently blocked for the observed neutron star masses to be allowed.

In \cite{Pfutzner:2018ieu} it has been argued that the unexpectedly high production of $^{10}{\rm Be}$ in the decay of $^{11}{\rm Be}$  \cite{Riisager:2014gia} can be explained by the neutron dark decay proposed in this paper.

\bibliography{neutron}

\end{document}